\documentclass[a4paper,10pt,twocolumn,nobalancelastpage,aps,pra,superscriptaddress,nofootinbib,longbibliography%,linenumbers % <<<1
]{revtex4-1}

\usepackage[utf8]{inputenc}
\usepackage[T1]{fontenc}
\usepackage[kerning=true,protrusion=true,expansion=true]{microtype}

\usepackage{amssymb,amsmath,amsthm,yfonts,pifont,bm,mathrsfs,lettrine,tabu,siunitx,stmaryrd}
\usepackage{booktabs,longtable}

\usepackage{graphicx,color,framed,mdframed}
\usepackage[dvipsnames]{xcolor}
\usepackage[ttdefault]{AnonymousPro}
\graphicspath{{./fig/}{./figs/}{./tikz/}}

\usepackage{datetime}
\settimeformat{xxivtime}

\newdateformat{usvardate}{%
	\monthname[\THEMONTH] \THEDAY, \THEYEAR}
\usepackage{tikz}
\usetikzlibrary{3d, calc, arrows.meta, decorations.pathmorphing, automata, backgrounds, petri, positioning, fit, patterns, decorations.pathmorphing, decorations.markings, scopes, shapes, shapes.geometric, shapes.symbols, shapes.multipart, shapes.callouts, shapes.misc, intersections}
\tikzset{>=stealth}
\pgfdeclarelayer{deep}
\pgfdeclarelayer{deepest layer}
\pgfdeclarelayer{background layer}
\pgfdeclarelayer{foreground layer}
\pgfsetlayers{deepest layer,deep,background layer,main,foreground layer}
\newcommand{\onback}{\begin{pgfonlayer}{background layer}}
		\newcommand{\outback}{\end{pgfonlayer}}
\usepackage[pdftex,
	colorlinks,
	allcolors=blue,
	pdfauthor={A. Rakhubovsky},
	pdftitle={},
	pdfsubject={},
	pdfkeywords={},
	pdfproducer={Latex with hyperref},
	pdfcreator={pdflatex}]{hyperref}

\usepackage[capitalize,poorman]{cleveref}

\newcommand*{\s}[1]{\ensuremath{_\mathsf{#1}}}
\newcommand*{\up}[1]{\ensuremath{^\mathsf{#1}}}
\newcommand*{\dd}[1]{\ensuremath{\mathrm{d}#1\:}}
\newcommand*{\da}{\dagger}
\newcommand{\sym}[2]{\left\langle #1 \circ #2 \right\rangle}
\providecommand*{\Mode}[1]{\ensuremath{{\mathsf{#1}}}}
\providecommand*{\cff}[1]{\ensuremath{\mathcal{#1}}}
\newcommand*{\cG}{\cff{G}}
\newcommand*{\cT}{\cff{T}}

\newcommand*{\avg}[1]{\left\langle #1 \right\rangle}
\newcommand*{\mat}[1]{\ensuremath{\mathbb{#1}}}
\renewcommand*{\vec}[1]{\ensuremath{\textbf{#1}}}
\newcommand*{\comm}[2]{\ensuremath{ \left[ #1 , #2 \right] }}
\newcommand*{\acomm}[2]{\ensuremath{ \left\{ #1 , #2 \right\} }}
\newcommand*{\mel}[3]{\left\langle#1\middle|#2\middle|#3\right\rangle}

\DeclareMathOperator*{\Var}{\operatorname{Var}}
\DeclareMathOperator*{\diag}{\operatorname{diag}}

\DeclareMathAlphabet{\mathbbold}{U}{bbold}{m}{n}
\newcommand{\II}{\ensuremath{\mathbbold{1}}}

\begin{document}

\newcommand*{\UPOL}{\affiliation{Department of Optics, Palack{\'y} University, 17.~Listopadu~12, 771~46~Olomouc, Czech~Republic }}
\newcommand*{\UV}{\affiliation{Vienna Center for Quantum Science and Technology (VCQ), Faculty of~Physics, University of~Vienna, Boltzmanngasse~5, A-1090~Vienna, Austria}}
\newcommand*{\IQOQI}{\affiliation{Institute for Quantum Optics and Quantum Information (IQOQI), Boltzmanngasse 3, 1090 Vienna, Austria}}

\newcommand*{\thetitle}{Detecting nonclassical correlations in levitated cavity optomechanics}
\title{\thetitle}
\author{Andrey A. \surname{Rakhubovsky}}
\email[Corresponding author: ]{andrey.rakhubovsky@gmail.com}
\UPOL
\author{Darren W. \surname{Moore}}
\UPOL
\author{Uro\v s \surname{Deli\' c}}
\UV
\IQOQI
\author{Nikolai \surname{Kiesel}}
\UV
\author{Markus \surname{Aspelmeyer}}
\UV
\IQOQI
\author{Radim \surname{Filip}}
\UPOL

\begin{abstract}
	Nonclassical optomechanical correlations enable optical control of mechanical motion beyond the limitations of classical driving.
    Here we investigate the feasibility of using pulsed cavity-optomechanics to create and verify nonclassical phase-sensitive correlations between light and the motion of a levitated nanoparticle in a realistic scenario.
    We show that optomechanical two-mode squeezing can persist even at the elevated temperatures of state-of-the-art experimental setups.
We introduce a detection scheme based on optical homodyning that allows revealing nonclassical correlations without full optomechanical state tomography.
    We provide an analytical treatment using the rotating wave approximation~(RWA) in the resolved-sideband regime and prove its validity with a full numerical solution of the Lyapunov equation beyond the RWA.
    We build on parameters of current experiments for our analysis and conclude that the observation of nonclassical correlations is possible today.
\end{abstract}

\maketitle

\section{Introduction} % <<<1
\label{sec:introduction}

Nonclassical correlations between different quantum systems constitute the physical backbone of hybrid quantum devices. For example, electromagnetic radiation is used to connect atomic or solid-state quantum systems over long distances. In hybrid quantum systems, these interactions need to go beyond the limitations of classical interfaces. This can be verified by detecting nonclassical correlations, for example, using the Cauchy-Schwarz inequality for the second order correlation function~\cite{riedinger_non-classical_2016,seri_quantum_2017,dou_broadband_2018,dou_direct_2018} or by direct observation of energy correlations~\cite{iskhakov_generation_2009,furst_quantum_2011,finger_raman-free_2015-1,wei_twin-beam-enhanced_2019-1}.
These witnesses are phase-insensitive. In contrast, phase-sensitive witnesses rely on the detection of quadrature correlations between homodyne detectors~\cite{lvovsky_squeezed_2015,schnabel_squeezed_2017}.
Revealing such correlations is a precursor for the evaluation of continuous-variable entanglement~\cite{laurat_entanglement_2005,weedbrook_gaussian_2012} that builds up in the interaction between quantum oscillators and light. Experimental demonstrations of entanglement, or two-mode squeezing, have been achieved between light and atoms~\cite{josse_continuous_2004,blinov_observation_2004,sherson_quantum_2006-1} and between microwaves and mechanical motion~\cite{palomaki_entangling_2013, palomaki_coherent_2013}.
In the optical domain, phase-insensitive detection of nonclassical  correlations has been achieved in optomechanical photonic crystals~\cite{riedinger_non-classical_2016,hong_hanbury_2017,riedinger_remote_2018}.

The observation of nonclassical effects in cavity-optomechanics is often inhibited by mechanical coupling to a thermal environment.
It is therefore not surprising that most quantum optomechanics experiments to date are performed in a cryogenic environment.
Several strategies have been proposed to circumvent the negative impact of the thermal environment~\cite{vitali_optomechanical_2007,genes_robust_2008}.
One of the most promising approaches relies on utilizing pulsed cavity-optomechanical control~\cite{hofer_quantum_2011}, which was shown to be robust to the thermal noise of the mechanical environment and therefore does not rely on deep precooling of the mechanical motion~\cite{rakhubovsky_robust_2015}.
In experimentally implementing such an approach, levitated cavity-optomechanics~\cite{rashid_experimental_2016,goldwater_levitated_2018,delic_cavity_2019,windey_cavity-based_2019,meyer_resolved-sideband_2019,delic_cooling_2020} is particularly suitable due to its simple mechanical mode structure. Additionally, it provides an excellent isolation from the thermal environment~\cite{chang_cavity_2010,romero-isart_optically_2011}. With the recent demonstration of ground-state cooling in an optical cavity~\cite{delic_cooling_2020}, the demonstration of nonclassical optomechanical correlations constitutes a next relevant step for levitated optomechanics.

Here we demonstrate how phase-sensitive non-classical correlations can be observed in a state-of-the-art optical levitation experiment at room-temperature.
We analyze the entanglement and generalized squeezing that is created during a pulsed entanglement protocol with a moderate precooling (to an occupation of $10^{4}$ phonons) and realistic heating rates.
Our analysis builds on previous studies of the pulsed approach to optomechanical entanglement~\cite{hofer_quantum_2011,palomaki_entangling_2013}.
In contrast to these studies we also consider the non-adiabatic regime of a good cavity, in which the timescales of coupling and pulses exceed the cavity decay rate.
Moreover, we propose a setting that allows direct verification of the nonclassical correlations by detecting the relevant quadratures in homodyne detection. This approach does not require full quantum state tomography of the optomechanical state and thus requires less experimental effort.
We conclude that nonclassical correlations are observable in state-of-the-art levitated cavity-optomechanics in a room-temperature environment and without the need for pre-cooling to the ground state.

\section{Results} % <<<1
\label{sec:results}

\subsection{Generation of the nonclassical correlations} % <<<2
\label{sec:generation_of_the_nonclassical_correlations}

In this manuscript we consider a pulsed optomechanical protocol to create nonclassical phase-sensitive correlations between a pulse of radiation and a noisy mechanical oscillations of levitated nanosphere.
In levitated optomechanics, the radiation pressure enables parametric interaction between cavity light and oscillations of a nanoparticle trapped inside the cavity in an externally applied Gaussian beam.
The interaction can be enhanced and manipulated by a laser drive.
Depending on the tuning of the drive, different types of optomechanical interaction can be observed.
They can differ in their sensitivity to noise of mechanical systems~\cite{filip_robust_2013}.
When the optomechanical system operates in the so-called resolved-sideband regime (when the cavity linewidth $\kappa$ is smaller than the frequency of mechanical motion of the particle $\Omega$), in the particular case of the resonant drive on the upper mechanical sideband, an effective amplifier-type interaction is established between the light and mechanics, with the Hamiltonian expressed in terms of the ladder operators as $H\s{TMS} = g ( a_m a_L + \text{h.c.} )$.
This interaction is known to generate various forms of quantum correlations between electric field of light and the particle motion.
In Ref.~\cite{palomaki_entangling_2013} this type of coupling was used to create electromechanical entanglement after pre-cooling the mechanical oscillator from the initial occupation of approximately $n_0 = 40$ down close to the ground state.
Advantageously, the generated highly mixed entanglement can be quite robust against the mechanical noise for a highly developed experimental platform~\cite{rakhubovsky_robust_2015} even without this pre-cooling stage.
This property is an important resource for transducers with hot mechanical systems~\cite{bagci_optical_2014,andrews_bidirectional_2014,higginbotham_electro-optic_2018,vostrosablin_pulsed_2017,*vostrosablin_quantum_2018}.
Driving the system on the lower mechanical sideband enables parametric conversion between the light and mechanical motion with effective Hamiltonian $H\s{BS} = g (a_m a_L^\da + \text{h.c.})$.
Using the toolbox of these interactions, it is possible to create and verify the non-classical correlations in an optomechanical system.

\begin{figure}[htb!]
	\centering
	\includegraphics[width=\linewidth]{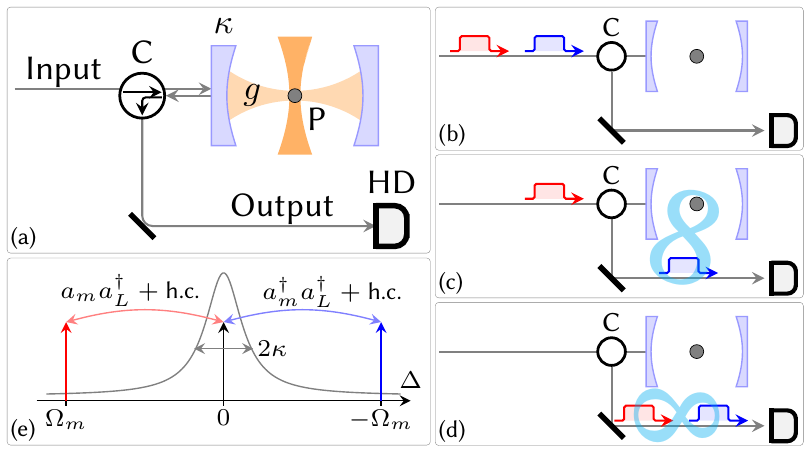}
	\caption{A scheme of the pulsed protocol to generate and verify nonclassical correlations between the levitated nanoparticle and light.
		(a) A sketch of the proposed setup.
		A subwavelength particle (P), trapped within a single-mode cavity, is coupled to its mode via radiation pressure at rate $g$. (b-d) Two light pulses, blue- and red-detuned enter the cavity to interact with the particle and are subsequently routed to the homodyne detector (HD) via the circulator (C). The cyan $8$-like symbol denotes non-classical correlations between different parts of the system. (e) Depending on the detuning $\Delta = \omega\s{cav} - \omega\s{p}$ of the pump, different types of interaction take place between the center-of-mass motion of the particle and the intracavity field. }
	\label{fig:scheme}
\end{figure}

The specific experimental setup we investigate is shown in~\cref{fig:scheme} (see also Refs.~\cite{delic_levitated_2020,delic_cooling_2020}) with all the relevant parameters that are used throughout the paper summarized in the table of~\cref{sec:parameters_used_for_simulations}.
An optical tweezer ($\lambda=1064$~nm) traps a silica nanoparticle (Radius: $r\approx 70$~nm) providing an essentially harmonic trap with a radial frequency of $\Omega_m = 2 \pi \times 190$~kHz.
The nanoparticle is positioned at the axis of a high-Finesse cavity (cavity linewidth $\kappa\approx 2 \pi \times 96.5$~kHz, Length: $l \approx 10$~mm) and at the intensity slope of the standing wave in an optically driven mode, resulting in the typical linearized optomechanical interaction (\cite{chang_cavity_2010,romero-isart_toward_2010}).
The particle is also interacting with a thermal environment given by the surrounding gas at a pressure of $10^{-6}$~mbar and room temperature resulting in a dominant heating rate of $15$~kHz.
Recoil from scattered photons adds a smaller contribution of $6$~kHz.
Following the protocol of~\cite{hofer_quantum_2011}, non-classical correlations are created by sending a blue-detuned light pulse (Detuning: $\Delta = \Omega_m$, Duration: $\tau=6.6$ $\mu$s~$=8/\kappa$, coupling to mechanics: $g=2\pi\times 60$~kHz) onto the cavity.
We assume the driving laser to be shot-noise limited.
Laser phase noise manifests itself in an additional heating of the particle's motion at a moderate rate and is therefore covered in the following analysis.
Precooling assumes only a moderate initial occupation of $n_0 \leq 10^4$ phonons, because no significant improvement of entanglement is achieved by lower initial occupations.
In~\cref{sec:generation_of_the_nonclassical_correlations} we analyze these optical-mechanical correlations in detail.
To detect the correlations a second, red-detuned readout pulse is used (detuning: $\Delta = -\Omega_m$, coupling $g$ and duration $\tau$ equal for both pulses).
We analyze the non-classical correlations observed between these pulses in homodyne detection after emission from the optical cavity in~\cref{sec:detection_of_the_correlations_with_homodyne_measurement}.
Losses of the cavity mirrors provide a dominant limitation on efficiency.
In our results we use current and realistic detection efficiencies of $0.4$ and $0.8$, respectively.

We propose to utilize this interaction of the center-of-mass displacement of a levitated nanoparticle with a pulse of light impinging upon the cavity to prepare a nonclassical correlation between the leaking light and noisy mechanical motion.
In the presence of the pump, the annihilation operators of the optical ($\Mode B$) and the mechanical ($a_m$) modes transform according to the set of linear input-output relations
\begin{subequations}
	\label{eq:ioblue:boson}
	\begin{align}
		\Mode B\up{out} & = \sqrt{ \cG } \Mode B\up{in} + \sqrt{ \cG - 1 } a_m^\da (0),
		\\
		a_m (\tau)      & = \sqrt{ \cG } a_m (0) + \sqrt{ \cG - 1 } \Mode B\up{in,}{}^\da,
	\end{align}
\end{subequations}
which corresponds to a two-mode squeezing interaction with the gain $\cG \approx \exp[ 2 g^2 \tau / \kappa]$ determined by the values of the optomechanical coupling rate $g$, cavity linewidth $\kappa$ and the interaction duration $\tau$.
It is important to stress, that $\Mode B\up{out}$ describes a certain mode of light defined by a particular temporal profile (see~\cref{sec:hamiltonian_formulation} for details).

\begin{figure}[t]
	\centering
	\includegraphics[width = \linewidth]{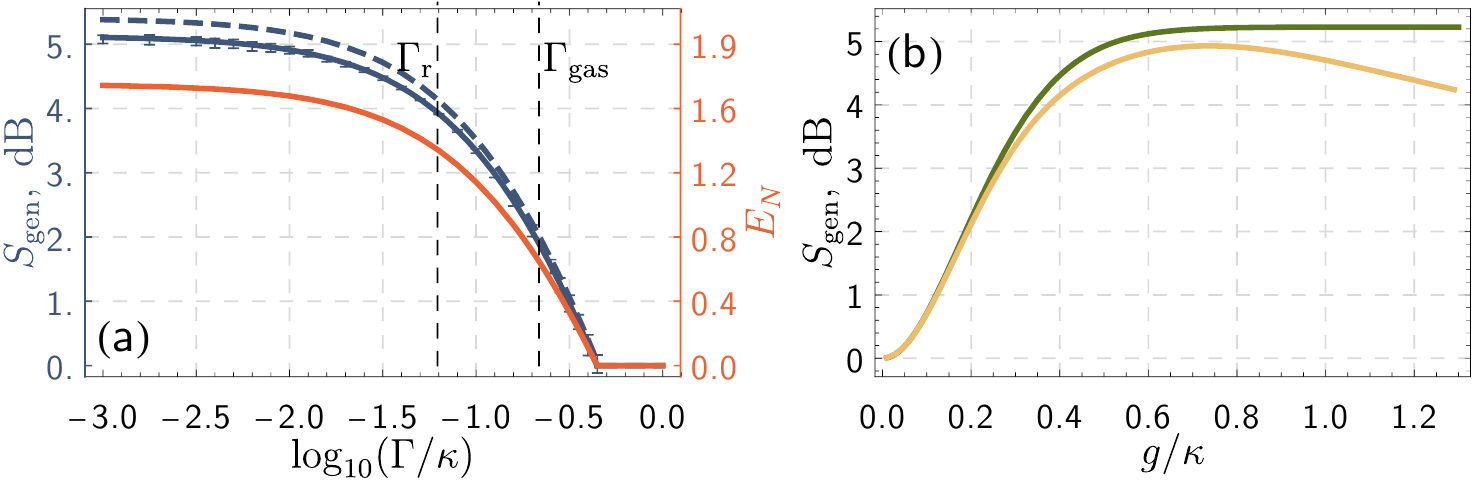}
	\caption{
		(a) Robustness of the nonclassicality of the correlation between light and mechanical motion of levitating nanoparticle (see~\cref{fig:scheme}~(c)).
		Two-mode squeezing $S\s{gen}$ (dark blue) and logarithmic negativity $E_N$ (light red) as functions of the heating rate $\Gamma$ (logarithmic scale, in units of $\kappa$) assuming initial mechanical occupation of $n_0 = 10^4$.
		The dashed line shows the two-mode squeezing computed without the rotating wave approximation.
		The error bars on the blue curve are obtained from the numerical simulation (see text).
		(b) Two-mode squeezing as a function of the coupling rate.
		The dark green line shows the adiabatic regime; light yellow, full solution.
		The two solutions coincide well at values corresponding to the weak coupling where the adiabatic approximation is valid.
		Numerical parameters are $g = 0.6 \kappa $, $\tau = 8 / \kappa$, $\Omega = 2 \kappa $.
	}
	\label{fig:robustness}
\end{figure}

It is known that the unitary two-mode squeezing interaction can entangle the participating modes even in the case of large initial noise in one of the modes~\cite{filip_robust_2013}.
The optomechanical cavity, however, is open and therefore its dynamics is not simply unitary.
The cavity is coupled to the detection channel and therefore has a finite linewidth~$\kappa$.
The collection efficiency is non-unity, so a part of the signal from the optomechanical cavity is lost.
This effect is easy to describe as a pure loss before the detection.
The nanoparticle experiences heating at rate $\Gamma$ due to a number of factors, of which the two most important ones are collisions with the residual gas inside the vacuum chamber and the recoil heating.
Our task is therefore to analyze the robustness of the protocol to these imperfections and determine to what extent the performance of a realistic levitated system can approach the idealised unitary two-mode squeezing.
The analysis of the operation of a levitated system has a few peculiarities compared to the conventional bulk devices.
The levitated systems are capable of demonstrating rather strong optomechanical coupling comparable to the cavity linewidth $g \sim \kappa$.
This allows operation beyond the adiabatic regime (in which the intracavity light mode is unpopulated).
This, in turn, requires a precise definition of the temporal modes for the homodyne detection.
Another peculiarity is a rather stringent limit on the duration of operation $\tau$ set by the inverse heating rate: $\Gamma \tau \leq 1$.

To visualize the nonclassical correlations in the system formed by the leaking pulse and the mechanical oscillator, we use entanglement evaluated in terms of logarithmic negativity~\cite{laurat_entanglement_2005} and the two-mode squeezing.
The latter has the meaning of the maximal amount of squeezing that can be directly extracted from the system by a global Gaussian passive transformation~\cite{weedbrook_gaussian_2012}.
Both quantities can be straightforwardly computed for a bipartite Gaussian quantum state.
We evaluate the nonclassicality as a function of the heating rate in~\cref{fig:robustness}(a).
It can be seen that the heating influences the correlations only beyond a certain point (in~\cref{fig:robustness}: approximately $10^{-2}$), after which the nonclassicality decreases and eventually vanishes after a certain value $\Gamma\s{crit}$.
In~\cref{fig:robustness} two values of the heating rate are emphasized: $\Gamma\s{gas}$ set by the collisions with the residual gas in the trapping chamber and the recoil heating~\cite{jain_direct_2016} and $\Gamma\s{r}$ set by the recoil heating alone.
The latter is determined by the properties of the particle and the trap, and appears to be a fundamental limitation for a given experimental environment.
We, however, see that this limitation is not prohibitive for the feasible setup.

The critical value of the heating rate depends on the particular regime of operation determined by the coupling rate $g$ and the pulse duration $\tau$.
For~\cref{fig:robustness} we numerically optimized the two-mode squeezing as a function of $g,\tau$ to persist at as high $\Gamma$ as possible.
One regime of interest is the so-called \emph{adiabatic} regime, in which the cavity mode is eliminated and the interaction approaches the ideal unitary form~\eqref{eq:ioblue:boson} of the pure two-mode squeezing.
If the cavity linewidth $\kappa$ is the dominant rate of the system, the cavity excitations instantaneously leak.
Thereby, the propagating light is virtually directly coupled to the mechanical motion, which allows transformations of the form~\eqref{eq:ioblue:boson}.
The formal requirements of this regime can be cast as $\kappa \geqslant g , \tau^{-1}, \Gamma$, therefore, the regime needs the pulses to be rather long with $\kappa \tau \gg 1$.
The available duration of the pulse, however, is limited from above by the heating rate.
Indeed, the decoherence caused by interaction with the environment is manifested in the form of an admixture of thermal noise to the quantum state of the particle with the variance of the thermal noise proportional to $\Gamma \tau$.
When the condition $\Gamma \tau \ll 1 $ is met the environment has virtually no impact on the nonclassicality which can be seen from~\cref{fig:robustness}(a).
This, however, with the currently observable heating rates, sets an upper limit on allowed pulse durations $\tau$.
This, in turn, makes the adiabatic regime infeasible, and makes the full dynamics richer than just unitary two-mode squeezing.
We, however, see from~\cref{fig:robustness}(b) that for weak coupling rates $g \ll \kappa$ that satisfy the adiabatic requirements, the performance of the full solution can be approximated by the adiabatic one.
Creation of strong nonclassical correlations, however, requires an excursion outside of the adiabatic regime.

To produce the curves we also assume that both the subsystems experience a loss characterized by transmittance $\eta = 0.8$ ($\eta = 1$ corresponding to the lossless case).
For the optical mode this loss describes different possible imperfections including non-perfect detection and collection efficiency, and also imperfect temporal and spatial mode overlap.
A readout of the mechanical oscillator's state done by a red-detuned pulse is also equivalent to an almost noiseless lossy map to light~\cite{filip_transfer_2015,rakhubovsky_photon-phonon-photon_2017}, so the lossy model gives a good grasp of this process.
Importantly, the figure shows that the nonclassical correlations of interest are quite robust to pure loss.
The correlations are also created regardless of the initial temperature of the mechanics.
\cref{fig:robustness}(a) is plotted for a realistic value of initial occupation of $n_0 = 10^4$, and decreasing the occupation further does not increase magnitudes of either the two-mode squeezing or the logarithmic negativity.

The error bars in~\cref{fig:robustness}(a) are obtained from numerical simulation.
A covariance matrix of the final bipartite state depends on a number of parameters, including the coupling rate $g$, interaction duration $\tau$, the initial occupation $n_0$ and the reheating rate $\Gamma$.
For a certain value of the latter we take an ensemble of 40 numerical samples of covariance matrices assuming that each of the other parameters $\epsilon$ (with $\epsilon$ being one of $g$, $\tau$ or $\log_{10} n_0$) is uniformly distributed within the region $0.9 \epsilon \leq \epsilon \leq 1.1 \epsilon$.

To perform a complete check of the robustness of our proposed setup, we also compute the two-mode squeezing beyond the rotating wave approximation by numerically solving the Lyapunov equation for the covariance matrix (see~\cref{sec:lyapunov_equation}).
This effectively includes the counter-rotating coupling terms that represent the beamsplitter-type interaction between light and mechanics.
Our analysis (\cref{fig:robustness}(a), dashed line) shows that for a moderate sideband resolution ($\Omega = 2 \kappa$) inclusion of these terms slightly improves the creation of nonclassical correlations thanks to additional cooling of the nanoparticle.
As $\Omega$ increases, the dashed line approaches the solid line until at approximately $\Omega = 10 \kappa$ there is no difference caused by inclusion of the beyond-RWA terms.
For details see~\cref{sec:lyapunov_equation}.

\subsection{Detection of the correlations via homodyne measurement} % <<<2
\label{sec:detection_of_the_correlations_with_homodyne_measurement}

In this section we provide a recipe to experimentally assess the nonclassical correlations between the light and mechanics created by the blue-detuned pulse.
The quantum correlations can be detected by swapping the mechanical state to the leaking pulse followed by a tomography of the quantum state of the two pulses.
Using homodyne detection one can reconstruct the full covariance matrix from which it is possible to evaluate an arbitrary measure of nonclassicality, in particular, the two-mode squeezing and the entanglement.
We also can design an experimentally less resource-demanding two-mode squeezing witness that can show the presence of entanglement requiring fewer experimental runs.
In particular we notice that the optomechanical entanglement manifests itself in the form of two-mode squeezing that is a squeezing of a weighted combination of quadratures of light and mechanics.
The combination quantitatively coincides with the one provided by the theory of the unitary two-mode squeezing in the adiabatic regime.

Our proposal of the evaluation of the correlations created by the blue-detuned pules relies on use of a subsequent pulse tuned to the lower mechanical sideband of the cavity.
In such a case, depending on the parameters, the interaction between the light and the nanoparticle's motion approaches the unitary beamsplitter-type coupling.
The input-output relations for this coupling, similarly to~\cref{eq:ioblue:boson}, can be cast in the form:
\begin{subequations}
	\label{eq:iored:boson}
	\begin{align}
		\Mode R\up{out} & = \sqrt{ \cT } \Mode R\up{in} + \sqrt{ \cT - 1 } a_m (\tau_i),
		\\
		a_m (\tau_f)    & = \sqrt{ \cT } a_m (\tau_i) + \sqrt{ \cT - 1 } \Mode R\up{in},
	\end{align}
\end{subequations}
where $\tau_i = \tau + \tau_D$ is the instant at which the state swap starts, this instant incorporates some delay $\tau_D$ after the entangling interaction has ended, and $\tau_f = \tau_i + \tau_R$ is the instant at which the state swap is completed.
The transmittance $\cT$ is determined by the strength of the pump and the duration of the swap $\tau_R$.
It is easy to show (see~\cref{sec:estimating_covariance_matrices}) that $a_m (\tau_i)$ is equivalent to $a_m (\tau)$ entering~\cref{eq:ioblue:boson} up to some thermal noise defined by the duration of the delay $\tau_D$.
The correlations between the quadratures of the modes defined by annihilation operators $\Mode B\up{out}$ and $\Mode R\up{out}$ will then reveal the original correlations between $\Mode B\up{out}$ and $a_m (\tau)$ created by the blue-detuned pulse.

\begin{figure}[t]%<<<
	\centering
	\includegraphics[width = \linewidth]{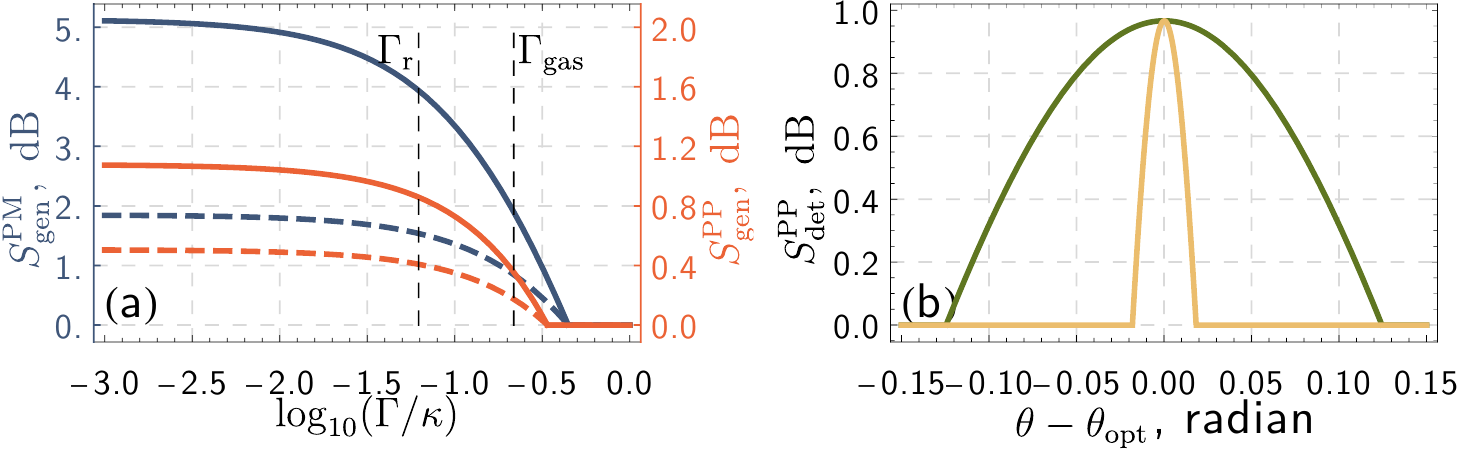}
	\caption{
		Detection of nonclassical correlations with homodyne measurement on leaking light.
		(a) comparison of the two-mode squeezing of the leaking entangling pulse and mechanical motion ($S\up{PM}$; dark blue; \cref{fig:scheme}~(c)) and two-mode squeezing of two pulses ($S\up{PP}$, light red; \cref{fig:scheme}~(d)) as a function of heating rate.
		Full and dashed lines show cases of detection efficiency $\eta = 0.8$ and $\eta = 0.4$ correspondingly.
		(b) possible detected two-mode squeezing $ - 10 \log_{10} \Var X\s{gen} [\phi\up{opt} , \theta\s{B} , \theta_R\up{opt}]$ as a function of a homodyne angle $\theta$ for two other angles set to their optimal values.
		Heating rate $\Gamma = 0.1 \kappa$.
		Different colors show initial occupation of mechanics $n_0 = 1$ for dark green and $n_0 = 100$ for light yellow.
	}
	\label{fig:eigenval}
\end{figure}%>>>

Using the standard approach outlined in the~\cref{sec:estimating_covariance_matrices} we can evaluate the nonclassical correlations that persist in the bipartite system formed by the two pulses.
\cref{fig:eigenval}, where we plot the two-mode squeezing as a function of the heating rate of mechanics, shows the robustness of the correlations to the mechanical decoherence.
It can be seen that compared to the purely unitary description of~\cref{eq:iored:boson} the readout is not purely lossy, but in fact adds noise.
The noise, however, is moderate so it does not prohibit creation of correlations between the two pulses with experimentally feasible parameters.
The graphs were obtained for the parameters optimized to make the correlations observable at as high heating rates $\Gamma$ as possible.
From the figure we conclude that with feasible heating rate and detection efficiency one can observe the two-mode squeezing in a state-of-the-art experiment.
The correlations between the two pulses are similarly robust to the initial mechanical occupation just as were the correlations between the mechanics and the entangling pulse.

A full tomography of the bipartite Gaussian state of the two pulses requires estimation of all 10 independent entries of their covariance matrix.
Instead, to reduce the experimental cost of the detection, we propose to use the fact that the two-mode squeezing manifests itself in the form of suppression of one of the eigenvalues of the CM below the level of shot noise.
This means that in the optimal basis, in which the CM is brought to the diagonal form, the smallest eigenvalue enters the CM as a variance of one of the quadratures.
Therefore, by adjusting the homodyne detection, it is possible to perform the measurement in this particular basis and to obtain the smallest eigenvalue directly.

To demonstrate the possibility of the two-mode squeezing observation we analyze a generalized quadrature of the bipartite system, defined as
\begin{equation}
	\label{eq:genquaddef}
	X\s{gen} [ \phi , \theta\s{B} , \theta\s{R} ] = X\s{B}^{\theta\s{B}} \cos \phi + X\s{R}^{\theta\s{R} }\sin \phi ,
\end{equation}
where we define the quadratures of individual subsystems in the bases rotated by an angle $\theta$: $X_i ^\theta :=  X_i \cos \theta + P_i \sin \theta$.
A certain choice of $\phi$ defines the weight with which quadratures of each of the systems enter the generalized quadrature.
In a particular experimental run, the values of angles $\theta$ are defined by the phase of local oscillator of the homodyne detector.
The angle $\phi$ can be adjusted in post-processing of the measurement results.
Transformation to the generalized quadrature~\eqref{eq:genquaddef} represents the most general passive Gaussian bipartite transformation possible.
The variance of $X\s{gen} [\phi\up{opt} , \theta\s{B}, \theta\up{opt}\s{R}]$ computed for our system is presented at~\cref{fig:eigenval}(b).
We observe that for certain values of $\phi\up{opt}$ and $\theta$ the generalized quadrature is squeezed below the shot noise level.
Both curves have the same magnitude, which shows the insensitivity of the correlations to the initial occupation $n_0$, however, a lower occupation simplifies the detection via homodyne since the squeezing can be seen in wider range of the homodyne angle $\theta$.
In other words, cooling closer to the ground state allows an order of magnitude less precise adjustment of the homodyne angles.
This is advantageous, since in an experiment the homodyne angles $\theta_\text{B,R}$ have to be set in advance to the values that have to be estimated from the theory.
Such an estimation relies on the calibration of the system parameters and can be error-prone.

\section{Conclusions} % <<<1
\label{sec:discussion}

In this manuscript we demonstrate the capability of linearized Gaussian dynamics of a levitated optomechanical system (a levitated particle inside an optical cavity) to create nonclassical correlations persisting at high initial temperatures of the mechanical mode.
Using the formalism of Heisenberg-Langevin equations and input-output theory we search for two-mode squeezing and entanglement (the logarithmic negativity).
While heating of the particle by the environment is the main limiting factor, we find that non-classical correlations can persist at parameters of current experiments.
These correlations are robust against experimental imperfections such as losses or heating effects, and can be efficiently detected by homodyning the cavity output field. This is achieved by a readout pulse that maps the correlations to a two-mode squeezed optical system.

The creation and observation of the described non-classical correlations is a powerful resource for quantum controlling levitated nanoparticles without the need to initialize the mechanical system in the quantum ground state. Future experiments may well combine such correlation based control with the additional capability of spatio-temporal driving of the optical potential landscape. These complementary methods can provide a rich toolbox for optimal non-Gaussian quantum state control.

\begin{acknowledgments}
	A.A.R, D.M. and R.F. were supported by the Czech Scientific Foundation (project 19-17765S), the MEYS of the Czech Republic (grant agreements No 731473 and 02.1.01/0.0/0.0/16\_026/0008460), the Czech Ministry of Education (project LTC17086 of INTER-EXCELLENCE program) and have received national funding from the MEYS and the funding from European Union’s Horizon 2020 (2014-2020) research and innovation framework programme under grant agreement No 731473 (project 8C18003 TheBlinQC). Project TheBlinQC has received funding from the QuantERA ERA-NET Cofund in Quantum Technologies implemented within the European Union’s Horizon 2020 Programme.
	U.D., N.K. and M.A. acknowledge support from the European Research Council (ERC CoG QLev4G), TheBlinQC (Project No. 864032; via the EC, the Austrian ministries BMDW and BMBWF and research promotion agency FFG), by the Austrian Science Fund (FWF): START programme (Y 952-N36) and doctoral school CoQuS (Project W1210), and the Austrian Marshall Plan Foundation.
\end{acknowledgments}

% >>>1

\appendix

\section{Parameters used for simulations} % <<<1
\label{sec:parameters_used_for_simulations}

\begin{table}[h!]
	\begin{tabular}{llll}
		Parameter            & Experimental          & Dimensionless & Simulations \\
		$\kappa$, kHz        & $ 2 \pi \times 96.5 $ & $1$           & 1           \\
		$g$, kHz             & $ 2 \pi \times 60$    & $0.62$        & $ \leq 0.6$ \\
		$\Omega_m$, kHz      & $ 2 \pi \times 190$   & $1.96$        & $2$         \\
		$\Gamma\s{gas}$, kHz & $ 2 \pi \times 15$    & $0.16$        & $0.16$      \\
		$\Gamma_r$, kHz      & $ 2 \pi \times 6$     & $0.06$        & $0.06$
	\end{tabular}
	\caption{Parameters of the experimental setup~\cite{delic_levitated_2020,delic_cooling_2020} used for simulations.}
	\label{tab:parameters}
\end{table}

The experimental parameters are listed as reported in~\cite{delic_levitated_2020,delic_cooling_2020}.
The reported experimental values are converted to dimensionless units where $\kappa = 1$ (third column).
In the fourth column are the values of the parameters used for the simulations to create~\cref{fig:robustness,fig:eigenval}.

\section{Estimating Covariance Matrices} % <<<1
\label{sec:estimating_covariance_matrices}

\subsection{Hamiltonian formulation} % <<<2
\label{sec:hamiltonian_formulation}

In this manuscript we consider a levitated optomechanical system as depicted in~\cref{fig:scheme}.
At the heart of the scheme is the levitated subwavelength dielectric particle trapped in high vacuum within a high-Q optical cavity assumed to be a single-mode cavity described by canonical quadratures $X_c, P_c$, with eigenfrequency $\omega\s{cav}$ and linewidth $\kappa$.
Within the cavity the subwavelength particle is held in a harmonic trap so that the particle's center of mass motion can be modelled as a harmonic oscillator with quadratures $X_m, P_m$ of eigenfrequency $\Omega$.
We normalize the quadratures such that $\comm{X_k}{P_k} = 2 i$, with $k = c,m$.
The two modes interact via radiation pressure by a linear coupling at rate $g$.
In the presence of a strong classical pump at~$\omega\s{p}$ the Hamiltonian of the system is linearized and reads in the frame rotating at~$\omega\s{p} a^\da a$ (see~\cite{law_interaction_1995,aspelmeyer_cavity_2014,bowen_quantum_2015}):
\begin{equation}
	\label{eq:apphamiltonian}
	H = \Delta ( X_c^2 + P_c^2)/4 + \Omega ( X_m^2 + P_m^2 )/4 - g X_c X_m,
\end{equation}
where $\Delta = \omega\s{cav} - \omega\s{p}$ is the detuning of the pump, $g = \frac{ \partial \omega\s{cav}}{ \partial X_m } \sqrt{\avg{ n\s{p}}}$ is the optomechanical coupling rate enhanced by  $\avg{n\s{p}}$, the mean number of intracavity photons due to the pump.

Taking into account damping and dissipation, one can write the system of Heisenberg-Langevin equations in the vector form
\begin{equation}
	\label{eq:langevin_vector}
	\dot{ \vec u } = \mat A . \vec u + \vec n,
\end{equation}
where $\vec u = (X_c , P_c , X_m , P_m)$ is the vector of unknowns, $\vec n = ( \sqrt{ 2 \kappa} X\up{in}, \sqrt{ 2 \kappa } Y\up{in} , 0 , \sqrt{ 2 \gamma } \xi\up{th})$ is the vector of input fluctuations.
We assume the optical input field to be in vacuum.
The mechanical fluctuations $\xi\up{th}$ stem from different sources including collisions with the residual gas particles and recoil heating.
We assume for $\xi\up{th}$ Markovian Gaussian statistics and describe it by a thermal state with the mean occupation $n\s{th}$, $\gamma$ is the viscous damping rate.
The two parameters are conveniently combined in the experimentally detectable reheating rate $\Gamma \equiv \gamma n\s{th}$.
To summarize, the noises obey the statistics (with the notation for the Jordan product $\sym{ \vec a }{ \vec b }_{ij } \equiv  \tfrac12 \avg{\Big( \vec a_i \vec b_j + \vec b_j \vec a_i \Big) }$):
\begin{gather}
	\sym{\xi\up{th}(t)}{\xi\up{th}(t')} = ( 2 n\s{th} + 1 ) \delta ( t - t' ),
	\\
	\sym{Q\up{in}(t)}{Q\up{in}(t')} = \delta ( t - t' ), \text{ with } Q = X,Y.
\end{gather}

The drift matrix $\mat A$ has the form
\begin{equation}
	\mat A =
	\begin{pmatrix}
		- \kappa & - \Delta & 0        & 0
		\\
		\Delta   & - \kappa & 2 g      & 0
		\\
		0        & 0        & 0        & \Omega
		\\
		2 g      & 0        & - \Omega & - \gamma
	\end{pmatrix}.
\end{equation}

\cref{eq:langevin_vector} has solution
\begin{equation}
	\label{eq:langevin_solutions}
	\vec u (t) = \mat M (t) . \vec u (0) + \int_0^t \dd s \mat M ( t - s ).
	\vec n (s),
\end{equation}
where $\mat M (t) \equiv \exp[ \mat A t ]$.
The solution is valid for a time-independent matrix $\mat A$.

One can write the solution for the light leaking from the cavity using input-output relations
\begin{equation}
	\label{eq:input_output}
	\vec u\up{out} = - \frac{ 1 }{ \sqrt{ 2 \kappa }}\tilde{ \vec n} + \sqrt{ 2 \kappa } \tilde{ \vec u },
\end{equation}
where tilde means taking the first two elements of a vector, e.g. $\tilde{ \vec u } = ( X_c ,  P_c )$.
The leaking field $\vec u\up{out} (t) = (X\up{out} (t) , Y\up{out} (t))$ describes a continuum of modes of semi-infinite space.
Of those we are interested in the one mode that is being detected in the homodyne detector.
The quantum state of the leaking pulse is therefore described by the canonical quadratures of that mode defined as
\begin{equation}
	\label{eq:mode_define}
	\vec U \equiv (\Mode X\up{out} , \Mode Y\up{out}) =  \int_0^\tau \dd s \vec u\up{out} (s) f\up{out} ( s ),
\end{equation}
where the function $f\up{out}$ defines the temporal profile of the mode to be measured.
In order to fully capture the necessary information it is important to tailor a specific profile.
This is particularly evident for the readout (red) pulse that necessarily needs to be adjusted to retrieve the state of mechanics at the beginning of this pulse since the initial mechanical state couples to a particular temporal mode.
For both pulses, we use the profiles equal to the element of the matrix $\mat M$ which corresponds to the transfer of the initial mechanical quadratures to the light.
In particular, for the readout pulse, we choose $f\up{out} (t) \propto \mat M_{13} (t)$.
Optimization of the profiles might be beneficial for the observation of stronger nonclassical correlation, though such observation is beyond the scope of the present work.
In an experiment, different profiles~$f\up{out}$ can be used by means of frequent sampling of the leaking signal and assembling a weighted sum of the samples in post-processing~\cite{morin_experimentally_2013,brecht_photon_2015}.
The state-of-the art detectors allow detection at rates that significantly exceed the characteristic rates of the levitated optomechanical systems, therefore there is a certain freedom in selecting~$f\up{out}$.

\subsection{Rotating Wave Approximation} % <<<2
\label{sec:rotating_wave_approximation}

In a typical optomechanical system, particularly in the levitated optomechanics, the optomechanical coupling rate is smaller than the cavity linewidth: $g \ll \kappa$.
Combined with the condition of the resolved sideband $\kappa \ll \Omega$ this suggests that creation of the optomechanical correlations happens at the timescale slower than the oscillations at $\Omega$.
This justifies a transition to the rotating frame, defined by the first two terms of~\cref{eq:apphamiltonian} $ H\s{rf} = \Delta ( X_c^2 + P_c^2)/4 + \Omega ( X_m^2 + P_m^2 )/4 ,$ with a subsequent application of rotating wave approximation (RWA).

The transition to the rotating frame amounts to the transformation from the vector $\vec u$ to the vector of the quadrature amplitudes $\vec u = \mat R. \vec v$, where $\mat R = \mat R_2 (\Delta t) \oplus \mat R_2 (\Omega t )$, and
\begin{equation}
	\mat R_2 (\alpha) =
	\begin{pmatrix}
		\cos \alpha   & \sin \alpha
		\\
		- \sin \alpha & \cos \alpha
	\end{pmatrix}.
\end{equation}
The new vector of unknowns contains the quadrature amplitudes $\vec v = ( X_c^{(c)}, X_c^{(s)}, X_m^{(c)}, X_m^{(s)} )$ that on the timescale of mechanical frequency do not change significantly.
The quadrature amplitudes obey a vector equation
\begin{equation}
	\label{eq:langevin_slow}
	\dot{ \vec v } = \mat A^{(v)} . \vec v + \vec n^{(v)},
\end{equation}
where
\begin{equation}
	\mat A^{(v)} = \mat R^{-1}. \left( \mat A \mat R - \dot { \mat R } \right),
	\quad
	\vec n^{(v)} = \mat R^{ - 1} . \vec n.
\end{equation}
The matrix $\mat A^{(v)}$ defined in this fashion contains terms that are rapidly oscillating compared to the rates $g, \kappa, \gamma$.
The RWA amounts in dropping these terms.
As a result we obtain
\begin{equation}
	\mat A_{\vec v } \rvert\s{RWA} =
	\begin{pmatrix}
		- \kappa  & 0         & 0                & g_r - g_b
		\\
		0         & - \kappa  & g_r + g_b        & 0
		\\
		0         & g_r - g_b & - \frac \gamma 2 & 0
		\\
		g_r + g_b & 0         & 0                & - \frac \gamma 2
	\end{pmatrix}.
\end{equation}
Here $g_r$ and $g_b$ are the optomechanical coupling rates enabled by pumping on the, correspondingly, lower and upper mechanical sidebands of the cavity.
We assume that during the entangling (blue-detuned, $\Delta = - \Omega$) pulse, $g_b = g , g_r = 0$.
In a similar fashion, during the read-out pulse (red-detuned, $\Delta = \Omega$), we put $g_b = 0, g_r = g$.
Between the pulses, there is no optomechanical interaction so the two modes (optical and mechanical) only expreience decay and decoherence.
The statistics of $\vec n^{(v)}$ takes simple form in RWA:
\begin{equation}
	\sym{ \vec n^{(v)} (t) }{ \vec n^{(v)} (t') } =  \diag[ 2 \kappa , 2 \kappa , 2 \Gamma , 2 \Gamma ] \delta (t - t').
\end{equation}

In the RWA then,~\cref{eq:langevin_slow} is a system of linear differential equations which can be solved in a manner similar to~\cref{eq:langevin_solutions}.
With help of~\cref{eq:input_output,eq:mode_define} where again $f\up{out} (t) \propto \mat M_{13} (t)$ one can write solution for the necessary covariance matrix.

% >>>
\subsection{Lyapunov Equation} % <<<2
\label{sec:lyapunov_equation}

One way to investigate the Gaussian dynamics described by~\cref{eq:langevin_slow} is to use Lyapunov equation.
For the covariance matrix $\mat V$ corresponding to the vector $\vec v$ it reads
\begin{equation}
	\dot{ \mat V } = \mat A^{(v)}. \mat V + \mat V. \left( \mat A^{(v)}\right)\up{T} + \mat D^{(v)},
\end{equation}
where we keep the terms oscillating at $2 \Omega$ in $\mat A^{(v)}$, and $\mat D^{(v)}$ is the diffusion matrix containing correlations of the noises:
\begin{equation}
	\sym{ \vec n^{(v)} (t) }{ \vec n^{(v)} (t') } = \mat D^{(v)} \delta ( t - t'),
\end{equation}
beyond RWA it reads
\begin{multline}
	\mat D^{(v)}
	\\
	=
	\begin{pmatrix}
		2 \kappa & 0        & 0                                    & 0                                         \\
		0        & 2 \kappa & 0                                    & 0                                         \\
		0        & 0        & 2 \Gamma ( 1 + \cos ( 2 \Omega t ) ) & 2 \Gamma \sin ( 2 \Omega t )              \\
		0        & 0        & 2 \Gamma  \sin ( 2 \Omega t )        & 2 \Gamma \sin ( 1 + \cos ( 2 \Omega t ) )
	\end{pmatrix}.
\end{multline}

We are, however, interested in the solution for the pulses of the leaking light.
Taking a derivative of~\eqref{eq:mode_define} over time, we can obtain equations of motion for the pulse quadratures.
Then we can write for a $6-$vector $\vec w = ( [\vec v]_{1 \times 4 } , \Mode X\up{out} , \Mode Y\up{out} )$
\begin{equation}
	\dot {\vec w } = \mat A^{(w)} \vec w + \vec n^{(w)},
\end{equation}
where
\begin{equation}
	\mat A^{(w)} =
	\begin{pmatrix}
		\left[ \mat A^{(w)} \right]_{4\times 4} & 0_{4\times 2}
		\\
		\begin{matrix}
			\sqrt { 2 \kappa } f\up{out} (t) \II_2 & 0 _{ 2\times 2 }
		\end{matrix}
		                                        &
		0 _{2 \times 2 }
	\end{pmatrix},
\end{equation}
with $\II_n$ and $0_{m \times n}$ being, respectively, an identity matrix and a matrix full of zeros of corresponding dimensions.
For the new $6 \times 6$ diffusion matrix we obtain
\begin{multline}
	\mat D^{(w)}
	\\
	=
	\begin{pmatrix}
		[ \mat D^{(v)} ]_{4 \times 4 }      &                  & - f\up{out} (t) \sqrt{ 2 \kappa } \II_2 \\
		                                    &                  & 0_{2 \times 2 }                         \\
		- f\up{out} \sqrt{ 2 \kappa } \II_2 & 0 _{2 \times 2 } & ( f\up{out} (t) ) ^2 2 \kappa \II_2
	\end{pmatrix}.
\end{multline}

In this definitions, the covariance matrix $\mat W = \sym{ \vec w }{ \vec w }$ can be obtained as a solution of the Lyapunov equation
\begin{equation}
	\dot{ \mat W } = \mat A^{(w)} . \mat W + \mat W . \mat A^{(w)}  + \mat D^{(w)},
\end{equation}
with the initial condition
\begin{equation}
	\mat W (0) = \diag ( 1 , 1 , 2 n_0 + 1 , 2 n_0 + 1, 0 , 0 ).
\end{equation}
The equation above shows that the intracavity mode starts at a vacuum state, the mechanics is initially in a thermal state with mean occupation $n_0$.

Removing the first two rows and two columns of $\mat W$ that correspond to the intracavity mode, we arrive to the relevant covariance matrix that describes correlations between mechanics and the pulse.
The Lyapunov equation can be further generalized in a fully similar fashion to describe the subsequent red-detuned readout pulse.
To do so, one has to extend the vector $\vec w$ to have the quadratures of the pulse, and properly generalize the matrices $\mat A^{(w)}$ and $\mat D^{(w)}$.

% >>>1
\section{Figures of nonclassicality} % <<<1
\label{sec:figures_of_nonclassicality}

The formalism of the Heisenberg-Langevin equations allows expression of the output operators $\vec r = (\Mode X\up{out} , \Mode Y\up{out} , X_m (\tau ) , Y_m (\tau) )$ in terms of linear combinations of the input ones.
The latter all have Gaussian statistics provided by thermal and vacuum states and therefore, since the dynamics of the system is described by linear differential equations, the output is in a Gaussian quantum state as well.
Owing to this, the output state of our system is completely described by the first two moments of the vector $\vec r$: the vector of means $\avg{\vec r}$ and the covariance matrix $\mat V$ with elements
\begin{equation}
	\mat V_{ij } = \frac 12 \avg{ \acomm{\vec r_i - \avg{ \vec r }_i }{\vec r_j - \avg{ \vec r }_j }  }.
\end{equation}
Here $\acomm{\cdot}{\cdot}$ denotes the anticommutator.
To evaluate the elements of $\mat V$ one requires the solution for $\vec r$ that is obtained as a combination of~\cref{eq:langevin_solutions,eq:input_output} and the knowledge of the statistics of the input states.
We assume the intracavity and the input optical mode to be in vacuum, the mechanical mode precooled to a thermal state with average occupation $n_0$ and the mechanical thermal noise in a thermal state with average occupation $n\s{th}$.

In this paper we evaluate the nonclassicality of the system using \emph{two-mode squeezing} and \emph{logarithmic negativity} as figures of merit.

The two-mode squeezing is shown by the smallest eigenvalue $\sigma$ of the covariance matrix of the  optomechanical system $\mat V$.
The two-mode squeezing has an intuitive interpretation as the squeezing that can be extracted from the system by a passive two-mode unitary operation.
A convenient way to evaluate the squeezing is to use the units of decibel (dB).
The conversion rule reads
\begin{equation}
	S\s{dB} = - 10 \log_{10} \frac{ \sigma }{ \sigma\s{vac}},
\end{equation}
where $\sigma$ is the squeezed eigenvalue.
Here $\sigma\s{vac}$ stands for the zero-point fluctuations variance ($\sigma\s{vac} = \mel{0}{ X^2 }{ 0} = 1$ for the choice of units in the present manuscript).

The logarithmic negativity of a bipartite quantum state shows the upper bound of distillable entanglement.
It can be computed~\cite{laurat_entanglement_2005,weedbrook_gaussian_2012} as
\begin{equation}
	E_N = \max[ 0 , - \log \nu_- ],
\end{equation}
where $\nu_-$ is the smaller symplectic eigenvalue of the covariance matrix of the partially transposed quantum state.
For the state with CM $\mat V$ written in block form as
\begin{equation}
	\mat V =
	\begin{bmatrix}
		\mat V_L       & \mat V_c \\
		\mat V_c\up{T} & \mat V_m
	\end{bmatrix}
\end{equation}
it reads
\begin{equation}
	\nu_- = \frac{ 1 }{ \sqrt{ 2 }} \sqrt{ \Sigma_V - \sqrt{ \Sigma_V ^2 - 4 \det \mat V }},
\end{equation}
with $\Sigma_V = \det V_m + \det V_L - 2 \det V_c$.

% >>>1

\bibliography{levitated_ascii}
\end{document}